\documentclass[showpacs,preprintnumbers,amsmath,amssymb,twocolumn]{revtex4}
\usepackage{amsmath}
\usepackage{graphicx}
\usepackage{epsf}
\usepackage{psfrag}
\usepackage{epsfig}
\usepackage{graphics}

\newcommand{\be}{\begin{equation}}
\newcommand{\ee}{\end{equation}}
\newcommand{\bq}{\begin{eqnarray}}
\newcommand{\eq}{\end{eqnarray}}

\begin{document}

\title{{Quantum gravitational contributions to the beta function of quantum electrodynamics}}
\date{\today}

\author{J. C. C. Felipe$^{(a)}$}\email[]{jean@fisica.ufmg.br}
\author{L. C. T. Brito$^{(b)}$} \email []{lcbrito@dex.ufla.br}
\author{Marcos Sampaio$^{(a)}$} \email []{msampaio@fisica.ufmg.br}
\author{M. C. Nemes$^{(a)}$}\email[]{mcnemes@fisica.ufmg.br}

\affiliation{(a) Universidade Federal de Minas Gerais - Departamento de F\'{\i}sica - ICEx \\ P.O. BOX 702,
30.161-970, Belo Horizonte MG - Brazil}
\affiliation{(b) Universidade Federal de Lavras - Departamento de Ci\^encias Exatas\\
P. O. BOX 3037, 37.200-000, Lavras MG - Brazil}

\begin{abstract}
We show in a diagrammatic and regularization independent analysis that the quadratic contribution to the beta function which has been conjectured to render quantum electrodynamics asymptotically free near the Planck scale has its origin in a surface term. Such surface term is intrinsically arbitrarily valued and it is argued to vanish in a consistent treatment of the model.
\noindent
\end{abstract}
\maketitle

Because of the negative mass dimension of the coupling constant perturbative  Einstein quantum gravity (EQG)
is nonrenormalizable \cite{'tHooft:1974bx,Deser:1974cz}. However one can still make sense of EQG  if it is interpreted as an
effective quantum field theory within a low energy expansion of a more fundamental theory. In an effective field theory all interactions compatible with its essential  symmetry content  are in principle allowed into the Lagrangian
\cite{Weinberg:1978kz} and thus it establishes a systematic framework to calculate quantum gravitational effects \cite{Donoghue:1993eb}.

This approach has been used to study the asymptotic
behavior at high energies of quantum field theories that incorporate the gravitational field.
Robinson and Wilczek suggest that the gravitational field improve the asymptotic freedom
of pure Yang-Mills near the Planck scale \cite{Robinson:2005fj}.
However, a similar calculation in the Maxwell-Einstein theory suggests that such conclusion is gauge dependent \cite{Pietrykowski:2006xy}. In a contribution \cite{Toms:2007sk} in which  the effective action is calculated in a
gauge-condition independent version of the background field method using dimensional regularization
it is argued that the gravitational field plays no role in the beta function of the Yang-Mills coupling.
Another calculation using  conventional diagrammatic
methods confirms this conclusion
\cite{Ebert:2007gf}.

In a recent publication, D. Toms \cite{Toms:2010vy} claimed that quadratic divergent contributions  
were responsible  to improve
asymptotic freedom of fine structure constant by quantum gravity effects by using  proper time cutoff regularization  
and effective action methods. However, the physical reality  of  the result in \cite{Toms:2010vy}  has been questioned 
\cite{Ellis:2010rw,Anber:2010uj}.

The purpose of this contribution is to shed light on the origin of such controversies using only a diagrammatic analysis. As an effective model EQG is intrinsically regularization dependent and consequently regularization becomes part of the model. We show however that the quadratic contributions to the beta function stem from ambiguous, arbitrarily valued, regularization dependent surface terms. We  present the  one loop  calculation
of the vacuum polarization tensor of the Maxwell-Einstein theory, both with and without matter, in the Feynman  and harmonic gauges for the photon and graviton, respectively.  We carry out  calculations such that  regularization ambiguities are isolated from  divergent integrals and compare with the results found in the literature showing explicitly the origin of the ambiguities. We evaluate arbitrary parameters in both cutoff and dimensional regularization. Finally we argue that such ambiguities can be fixed on physical grounds demanding  transversality of the vacuum polarization tensor in the limit of weak gravity. Our analysis is based on the point of view discussed by Jackiw in \cite{Jackiw}. He argues that it can happen that radiative
corrections can give rise to arbitrary finite quantities which must be fixed either by symmetries of the underlying theory and/or, just as for infinite
radiative corrections, by experimental data.

We start with the Maxwell-Einstein Lagrangian

\begin{equation}\label{action}
 S_{ME}= \int d^{4}x\sqrt{-g}\left[\frac{2}{\kappa^{2}}R
  -\frac{1}{4}g^{\alpha\mu}g^{\beta\nu}F_{\alpha\nu}F_{\mu\beta}
\right].
\end{equation}
 As usual, $F_{\mu\nu}$ is the
 electromagnetic field  strength tensor, $R$ the curvature scalar and $g$ the metric determinant.

The Feynman rules can be directly obtained from (\ref{action})
linearizing the metric around a Minkowski background metric $\eta^{\mu\nu}=(1,-1,-1,-1)$

\begin{equation}
g^{\mu\nu} = \eta^{\mu\nu} + \kappa h^{\mu\nu}.
\end{equation}
In the harmonic gauge, the graviton propagator reads

\begin{equation}\label{propagator_grav}
\Delta^{\alpha\lambda\sigma\beta}(p)=\frac{iP^{\alpha\lambda\sigma\beta}}{(p^{2}-\mu^{2}+i\epsilon)},
\end{equation}
with
\begin{equation}
P^{\alpha\lambda\sigma\beta}=\frac{1}{2}\Big(\eta^{\beta\lambda}\eta^{\sigma\alpha} + \eta^{\beta\alpha}\eta^{\lambda\sigma} - \eta^{\alpha\lambda}\eta^{\sigma\beta}\Big),
\end{equation}
while in the Feynman gauge the photon propagator is

\begin{equation}\label{foton_propag.}
\Delta^{\mu\nu}(p)=\frac{-i\eta^{\mu\nu}}{(p^{2}-\mu^{2}+i\epsilon)}.
\end{equation}
Here we introduce a infrared regulator $\mu$ which will be taken to zero in the end of the calculation.

In the diagram 1a, the trilinear vertex
\begin{figure}
\begin{center}\includegraphics{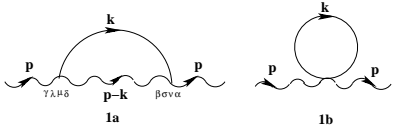}\end{center}
\caption{One-loop gravitational correction to the photon vacuum polarization in  Maxwell-Einstein theory.
Wavy lines are associated with the photon and  straight  lines with the  graviton.}
\label{feynman}
\end{figure}
can be translated into the Feynman rule

\begin{eqnarray}\label{vertice1}
& &\tau^{\lambda\theta\gamma\delta}(p,p^{\prime}) =  i\kappa\Big\{P^{\lambda\theta\gamma\delta}(p\cdot p')
+\nonumber\\
& &\frac{1}{2}\left[ \eta^{\gamma\delta}\left(p^{\theta}p'^{\lambda}+p^{\lambda}p'^{\theta}\right) \right.
+\eta^{\lambda\theta}p^{\delta}p'^{\gamma}
- \eta^{\lambda\delta}p'^{\gamma}p^{\theta}\nonumber\\
& &-\eta^{\lambda\gamma}p^{\delta}p'^{\theta}
-\left.\eta^{\theta\gamma}p^{\delta}p'^{\lambda}-\eta^{\theta\delta}p^{\lambda}p'^{\gamma}\right]\Big\}.
\end{eqnarray}
The tadpole diagram in figure   \ref{feynman}b  yields  a quadratic divergence which will exactly
cancel a quadratic divergence in diagram  \ref{feynman}a. We will return to this point when we
compute diagram \ref{feynman}a.

The one-loop contribution corresponding to the diagram in figure
 \ref{feynman}a is given by
\begin{eqnarray}\label{one_loop}
\Pi^{\mu\nu}(p)&=& -\kappa^{2}\int_k
\bigg[\frac{\eta_{\delta\alpha}P_{\gamma\lambda\beta\sigma}}{(k^{2}-\mu^{2})[(k-p)^{2}-\mu^{2}]}\nonumber\\
&\times&\tau^{\gamma\lambda\mu\delta}(p,p-k)\tau^{\beta\sigma\nu\alpha}(p-k,p)\bigg],
\end{eqnarray}
where in $\tau^{\mu\nu\rho\sigma}$ above the momenta flow towards the vertice of the Feynman diagram. 

We isolate the divergent content of the amplitude above as basic divergent integrals following \cite{IR}
as a convenient method to evaluate the extent to which the final result  depends
of a particular choice of  regularization. We begin by using  in (\ref{one_loop}) the identity

\begin{eqnarray}
\frac{1}{(k+p)^2 - \mu^2} &=& \frac{1}{k^2 - \mu^2} -  \nonumber\\
& & \frac{2k \cdot p + p^2}{(k^2 - \mu^2)[(k+p)^2 - \mu^2]}
\end{eqnarray}
in order to eliminate the external momentum $p$ from the basic divergent integrals which will be expressed as
\begin{equation}\label{ilog}
I_{log}(\mu^{2}) = \int_k \frac{1}{(k^{2} - \mu^{2})^2}
\end{equation}
and
\begin{equation}\label{iquad}
I_{quad}(\mu^{2}) = \int_k
 \frac{1}{(k^{2} - \mu^{2})}.
\end{equation}
We adopt the abbreviation $\int_k \equiv \int d^4k/(2 \pi)^4$. 

After some tensorial algebra,
the one-loop photon vacuum polarization (\ref{one_loop}) reads
\begin{eqnarray}\label{final_generic}
& &\Pi^{\mu\nu}_{grav}(p) =-\kappa^2 \bigg[\frac{5}{12}F(p^{2})\left(p^2 \eta^{\mu\nu} - p^{\mu}p^{\nu}\right)p^2
\nonumber\\
& & -  I_{quad}(\mu^2) \left(p^2\eta^{\mu\nu} - p^{\mu}p^{\nu}\right) + \Upsilon^{\mu\nu}_{1}\bigg].
\end{eqnarray}
The quadratic divergent term $I_{quad}(\mu^2)$ will be canceled by the tadpole diagram  in figure 
\ref{feynman}b whereas $F(p^{2})$ stands for
\begin{equation} \label{coeff_final}
F(p^{2}) =  I_{log}(\mu^{2})-\frac{i}{16\pi^2}\ln\left(-\frac{p^{2}}{\mu^{2}}\right).
\end{equation}
The apparent infrared divergence is eliminated by using the regularization independent identity \cite{IR}
$$
I_{log}(\mu^2)-I_{log}(\lambda^2)= - \frac{i}{16 \pi^2 } \ln \Big( \frac{\mu^2}{\lambda^2}\Big)\,\,\,\,, \lambda \neq 0,
$$
in which $\lambda$ plays the role of renormalization group constant. Thus
\begin{equation}
F(p^{2}) =  I_{log}(\lambda^{2})-\frac{i}{16\pi^2}\ln\left(-\frac{p^{2}}{\lambda^{2}}\right).
\end{equation}
 Finally the term expressed by $\Upsilon_1^{\mu \nu}$ reads
\begin{eqnarray}\label{cutoff}
& &\Upsilon_{1}^{\mu\nu}=\frac{c_{1}}{12}p^{2}\left(13p^{2}\eta^{\mu\nu}-20p^{\nu}p^{\mu}\right) -\nonumber\\
& &\left[\frac{c_{2}}{2} + \frac{8 c_{3}}{3}\, p^{2} \right]\left(\eta^{\mu\nu}p^{2}-p^{\mu}p^{\nu}\right).
\end{eqnarray}
The coefficients $c_i$ ($i = 1,2,3$) have origin in differences between divergent loop integrals (that is, integrals which are independent of external momenta) of the same degree of superficial divergence, namely
\begin{eqnarray}\label{surface}
&&c_{1}\eta^{\mu\nu}=\frac{1}{4}\eta^{\mu\nu}I_{log}(\mu^{2})-\int_k\frac{k^{\mu}k^{\nu}}{(k^{2}-\mu^{2})^{3}}\nonumber\\
&&c_{2}\eta^{\mu\nu}=\frac{1}{2}\eta^{\mu\nu}I_{quad}(\mu^{2})-\int_k\frac{k^{\mu}k^{\nu}}{(k^{2}-\mu^{2})^{2}}\nonumber\\
&&c_{3}\eta^{\{ \mu\nu} \eta^{\alpha\beta\}}=\frac{1}{24}\eta^{\{ \mu\nu} \eta^{\alpha\beta\}}I_{log}(\mu^{2})-\nonumber\\ && \int_k \frac{k^{\mu}k^{\nu}k^{\alpha}k^{\beta}}{(k^{2}-\mu^{2})^{4}}, \nonumber\\
\end{eqnarray}
with
\begin{equation}
\eta^{\{ \mu\nu} \eta^{\alpha\beta\}}= \eta^{\mu\nu}\eta^{\alpha\beta} + \eta^{\mu\alpha}\eta^{\nu\beta}+\eta^{\mu\beta}
\eta^{\nu\alpha}.
\end{equation}
One can show that (\ref{surface}) can be written as surface terms, namely
$$
c_1 \eta_{\mu \nu} = \int_k \frac{\partial}{\partial k^\mu}\Bigg(\frac{k_\nu}{(k^2-\mu^2)^2}\Bigg),
$$
$$
c_2 \eta_{\mu \nu} = \int_k \frac{\partial}{\partial k^\mu}\Bigg(\frac{k_\nu}{(k^2-\mu^2)}\Bigg),\,\,\, \mbox{and}
$$
$$
c_3\eta_{\{ \mu \nu}\eta_{\alpha \beta \}}=\int_k \frac{\partial}{\partial k^{\beta}}\Bigg(\frac{4 k_\mu k_\nu k_\alpha}{(k^2-\mu^2)^3}\Bigg).
$$
They are regularization dependent and thus undetermined in principle undetermined according to Jackiw's conjecture save if symmetries or experiments can fix such arbitrariness. It is easy to check that $c_i$ ($i=1, 2, 3$) evaluate to zero in dimensional regularization whereas in momentum cutoff
\begin{equation}c_{1}=\frac{i}{128\pi^{2}},\,\,\
c_{2}=-\frac{i\Lambda^{2}}{64\pi^{2}} \,\,\, \mbox{and}\,\,\,
c_{3}=\frac{5i}{2304\pi^{2}},
\label{ccutoff}
\end{equation}
with $\Lambda \rightarrow \infty$. It has been shown that setting such surface terms to zero amounts to allowing shifts in the integration variable in the Feynman amplitudes. Gauge invariance of Green's functions are automatically satisfied within perturbation theory by setting $c_i =0$ and their generalizations to higher loops. Moreover this leads to momentum routing invariance in the Feynman diagram.

To make contact with other results in the literature let us evaluate the expression (\ref{final_generic}) for  $\Pi^{\mu\nu}(p)$ in both dimensional and cutoff regularizations. For this purpose we use the followings straightforward result
\begin{equation}\label{ilog_dim}
I_{log}^{DReg}(\lambda^{2})=-\frac{i}{16\pi^{2}}\left[\frac{2}{d-4}+\ln\left(\frac{\lambda^{2}}{\bar{\mu}^{2}}\right)\right] +  \mathcal{O}(d-4)
\end{equation}
and, in momentum cutoff regularization,
\begin{equation}\label{ilog_cut}
I_{log}^{\Lambda}(\lambda^{2}) = -\frac{i}{16\pi^{2}}\left[1+\ln\left(\frac{\lambda^{2}}{\Lambda^{2}}\right)\right]
+\mathcal{O}\left(\frac{\lambda^{2}}{\Lambda^{2}}\right),
\end{equation}
recalling that $\Lambda\rightarrow \infty$ can play the r\^ole of effective upper energy limit.
Finally, using (\ref{ilog_dim}) and (\ref{ilog_cut}) in (\ref{final_generic}) yields
\begin{eqnarray}\label{dim_final_general_fey}
& &\Pi_{DReg}^{\mu\nu}(p)=\frac{5\kappa^{2}i}{192\pi^{2}}\bigg[\frac{2}{d-4}+
\ln\left(-\frac{p^{2}}{\bar{\mu}^{2}}\right)\bigg]\nonumber\\
& & \times \left(\eta^{\mu\nu}p^{2}-p^{\mu}p^{\nu}\right)p^{2};
\end{eqnarray}
whereas
\begin{eqnarray}\label{cuttof_final_general_fey}
& &\Pi_{\Lambda}^{\mu\nu}(p) = \frac{5\kappa^{2}i}{192\pi^{2}}\bigg\{
 \left[\frac{2}{9} +   \ln \left( -\frac{p^{2}}{\Lambda^{2}} \right) \right] \, p^2\nonumber\\
& &- \frac{3}{10}\,\Lambda^{2}\bigg\}\left(\eta^{\mu\nu}
p^{2}-p^{\mu}p^{\nu}\right)-\frac{13i}{1536}\kappa^{2}p^4\eta^{\mu \nu} + \nonumber \\ & &
 \frac{5i}{384} \kappa^2 p^2 p^\mu p^\nu.
\end{eqnarray}
Some comments are in order. Firstly the coefficient of $\Lambda^2$ is the same as the one obtained by D. Toms in \cite{Toms:2010vy} where it is claimed to contribute to asymptotic freedom of the structure constant near the Planck scale. Secondly the polarization tensor is not transverse in cutoff regularization whereas it is tranverse in dimensional regularization. And last but not least notice that the term $\Lambda^2$ in (\ref{cuttof_final_general_fey}) stem from the arbitrarily valued surface term $c_2$. For a renormalizable model such surface terms are completely fixed by gauge invariance. Consider the vacuum polarization tensor of QED evaluated in this
 framework \cite{IR} as an illustration. We have
\be
\Pi_{\mu \nu }^{QED}=\int_k
\mbox{tr} \left\{ \gamma _\mu S(k+p)\gamma _\nu
S(k)\right\} ,
\label{tpqed}
\ee
where $S(k)$ is the fermion propagator. It can be written as \cite{IR}
\bq
 \Pi_{\mu \nu }^{QED} &=& \tilde\Pi_{\mu \nu } +
4\Bigg[c_2 \eta_{\mu\nu} + \Big( \frac{c_3}{3}-c_1\Big)p^2 \eta_{\mu \nu} -\nonumber\\ 
& &\Big( c_1-\frac{2 c_3}{3}\Big)p_\mu p_\nu\Bigg]
\label{qedvp}
\eq
where
\bq
&& \tilde\Pi_{\mu \nu } = \frac{4}{3} \Big(
p^2g_{\mu
\nu }- p_\mu p_\nu \Big) \times
\nonumber \\
&& \Bigg[ I_{log} (m^2) - \frac i{(4\pi
)^2} \Bigg( \frac
13+ \nonumber \\ && \frac{(2m^2+ p^2)}{p^2}
F(p^2,m^2)\Bigg) \Bigg] \, ,
\eq
$F(p^2;m^2)$ is defined by
\be
F(p^2,m^2) = \int_0^1 dz \, \ln \Big[
\frac{p^2z(1-z)-m^2}{-m^2}\Big]
\label{ilogz0}
\ee
and the arbitrary parameters $c_i$'s are defined as before. Notice that in this case gauge invariance fixes their values as $c_1=c_2=c_3=0$, which is the result we would have obtained should we had evaluated these parameters in dimensional regularization. Moreover a second possibility also renders a transverse vacuum polarization tensor for QED, namely $c_2=0$ and $c_3=2 c_1$. It is clear that cutoff regularization using (\ref{ccutoff}) breaks gauge invariance in this case.

Back to the Maxwell-Einstein theory, we see that, on gauge invariance grounds, one claim that the result expressed by (\ref{dim_final_general_fey}) is the correct leaving no room for the quadratic contribution which is originated from the surface term $c_2$. However as an effective model, usually the regularization is part of the model and one could think of restoring gauge symmetry by adding finite counterterms to the original Lagrangian. Although this point of view seems to be satisfactory, we show that when we add matter to the gravitational and photon field, which is the model studied in \cite{Toms:2010vy}, a consistent analysis determines that there is no quadratic contributions to the beta function of the structure constant leading to an asymptotically free theory near the Planck scale. We use scalar quantum electrodynamics coupled to gravity for simplicity. For fermionic matter the conclusions, mutatis mutandis, are identical. To one loop order the only gravitational contribution to $\Pi^{\mu\nu}(p)$ is given by
(\ref{final_generic}). So the relevant terms in the action are obtained by adding to (\ref{action})
the  contributions corresponding to the lagrangian
\begin{eqnarray}
\mathcal{L} &=&  - \frac{\left(Z_{3}-1\right)}{4}F_{\mu\nu}F^{\mu\nu}+\frac{Z_{4}}{4}F_{\mu\nu}\Box F^{\mu\nu}
+\nonumber\\
& &Z_{2}\partial_{\mu}\phi^{*}\partial^{\mu}\phi - iZ_{1}eA_{\mu}\left(\phi^{*}\partial^{\mu}\phi-\phi\partial^{\mu}\phi^{*}\right)\nonumber\\
& &  +Z_{1}e^{2}A^{\mu}A_{\mu}\phi^{*}\phi,
\end{eqnarray}
with the correspondent counterterms $Z_{1}$, $Z_{2}$, $Z_{3}$ and $Z_{4}$. Here the Lorentz indices are
raised and lowered by  $\eta^{\mu\nu}$. The full one-loop photon vacuum polarization tensor take the form
\begin{eqnarray}
\Pi^{\mu\nu}(p) &=& -\Bigg[F(p^{2})\left(\frac{e^{2}}{3}+\frac{5\kappa^{2}}{12}p^{2}\right) +
\nonumber\\
& &i\left[\left(Z_{3}-1\right)+Z_{4}p^{2}\right]\Bigg] \left(\eta^{\mu\nu}p^{2}-p^{\mu}p^{\nu}\right) \nonumber\\
& & -\kappa^{2}\Upsilon_{1}^{\mu\nu}-4e^{2}\Upsilon_{2}^{\mu\nu}
\label{qed}
\end{eqnarray}

where
\begin{eqnarray}
\Upsilon_{2}^{\mu\nu}=c_{2}\eta^{\mu\nu}-\left(c_{1}-\frac{1}{6}c_{3}\right)\left(\eta^{\mu\nu}p^{2}+2p^{\mu}p^{\nu}\right)
\end{eqnarray}
and $\Upsilon_{1}^{\mu\nu}$ is given by (\ref{cutoff}). Recall that the quadratic contribution comes from the surface term $c_2$ contained in both $\Upsilon_1^{\mu \nu}$ and $\Upsilon_2^{\mu \nu}$. Just as in the case of pure QED, $c_2$ breaks gauge invariance in the matter sector of (\ref{qed}). Hence we must set it to zero in the matter sector on gauge invariance grounds or equivalently one has to use dimensional regularization which automatically evaluates such surface terms to zero.  For consistency with the limit where $\kappa \rightarrow 0$, {\it the $c_2$ term which would originate a quadratic contribution to the fine structure beta function rendering the theory asymptotically free does not exist}.

A final comment is in order. It is well known that a naive cutoff in the three or four momenta in the loop integral violates gauge invariance. However some variations of this method in conjunction with Pauli-Villars or proper time regularization have been used in effective field theories because it is advantageous to introduce an explicit cutoff in such models. The proper time approach introduced by Schwinger \cite{Schwinger} is not free of ambiguities. Consider for instance the quadratically divergent integrals discussed in \cite{Varin}:
$$A=\int_k \frac{k^2}{(k^2-m^2)^2}$$
and
$$ B= I_{quad}(m^2) + m^2 I_{log}(m^2).
$$
Using the proper time approach via the identity
$$
\frac{\Gamma(n)}{(k^2+m^2)^n} = \int_0^{\infty}\, d\tau \, \tau^{n-1} \, \mbox{e}^{-\tau(k^2+m^2)}
$$
 yields for the divergent structure of $A$ and $B$ the results
$$A=\frac{i}{8\pi^2}(\Lambda^2-m^2\ln \Lambda^2)$$
and
 $$B= \frac{i}{16\pi^2}(\Lambda^2-2m^2\ln \Lambda^2) $$
instead of the expected equality $A=B$. In the approach we have discussed here the equality $A=B$ is built in our framework \cite{IR}.

\end{document}